\documentclass[12pt,a4paper]{article} 

\usepackage{graphics,graphicx,placeins,afterpage,makecell,tablefootnote,
multicol,multirow,caption,textcomp,lineno,tabularx,subfigure,booktabs,
courier,collref,setspace,float}
\usepackage[utf8]{inputenc}
\usepackage[english]{babel}
\usepackage[backend=bibtex,style=authoryear,bibstyle=authoryear,maxcitenames=2,maxbibnames=7,useprefix=true]{biblatex}

\DeclareNameAlias{sortname}{last-first}

\begin{document}

\begin{center}
{\Large{Measurement of CH$_3$D on Titan at Submillimeter Wavelengths}}
\end{center}

\begin{center}
Authors: Alexander E. Thelen$^{a,b,}$\footnote{Corresponding
  author. (A. E. Thelen) Email
  address: alexander.e.thelen@nasa.gov.} 
, C. A. Nixon$^a$, M. A. Cordiner$^{a,b}$, S. B. Charnley$^a$,
P. G. J. Irwin$^c$, Z. Kisiel$^d$
\bigskip

{\footnotesize{$^a$NASA Goddard Space
    Flight Center, $^b$Catholic University of
America, $^c$University of Oxford, $^d$Polish Academy of Sciences}}
\end{center}
\bigskip

\begin{center}
(Received February 15, 2019; Revised March 27, 2019; Accepted April
16, 2019)
\end{center}

\begin{abstract}
We present the first radio/submillimeter detection of monodeuterated methane (CH$_3$D) in Titan's
atmosphere, using archival data from of the Atacama Large Millimeter/submillimeter Array (ALMA).
The $J_K=2_1-1_1$ and $J_K=2_0-1_0$ transitions at 465.235 and 465.250 GHz
($\sim0.644$ mm) were measured at significance levels of $4.6\sigma$
and $5.7\sigma$, respectively. These two
lines were modeled using the Non-linear optimal Estimator for
MultivariatE spectral analySIS (NEMESIS) radiative transfer code to determine the disk-averaged
CH$_3$D volume mixing ratio = $6.157\times10^{-6}$ in Titan's
stratosphere (at altitudes $\textgreater130$ km). By comparison with
the CH$_4$ vertical abundance profile measured by
\textit{Cassini-Huygens} mass spectrometry, the resulting value for
D/H in CH$_4$ is $(1.033\pm0.081)\times10^{-4}$. This is consistent
with previous ground-based and \textit{in-situ} measurements
from the \textit{Cassini/Huygens} mission, though slightly lower than
the average of the previous values. Additional CH$_3$D observations at
higher spatial resolution will be required to determine a value truly
comparable with the \textit{Cassini-Huygens} CH$_4$ measurements, by measuring
CH$_3$D with ALMA close to Titan's equator. In the
post-\textit{Cassini} era, spatially resolved observations of CH$_3$D
with ALMA will enable the latitudinal distribution of methane to be determined, making this an important
molecule for further studies. 

\end{abstract}

\section{Introduction} \label{sec:intro}
Methane (CH$_4$), one of the primary constituents of Titan's atmosphere, was
first detected by \textcite{kuiper_44} and has since been studied
through myriad ground- and space-based observations. The photo- and
ionic chemistry of CH$_4$ and
Titan's most abundant atmospheric constituent, N$_2$, in
the upper atmosphere are responsible for the large
quantity of trace organic species discovered during the
\textit{Voyager-1} and \textit{Cassini} eras, and are important for
the formation of haze, clouds, and Titan's methane-based hydrological
cycle. However, the source of Titan's atmospheric CH$_4$ reservoir is still
unknown, as are the processes by which it may be replenished. 

Titan's CH$_4$ volume mixing ratio was found to be
  $5.65\times10^{-2}$ near the surface at the location of the \textit{Huygens} landing site ($\sim10^{\circ}$
S) with the Gas
Chromatograph Mass Spectrometer (GCMS), decreasing with altitude to
the tropopause ($\sim45$ km); an average stratospheric value of $1.48\times10^{-2}$
was measured between $\sim75-140$ km \parencite{niemann_10}. This measurement
is compatible with the initial determination of stratospheric CH$_4$
abundance derived from the 7.7
$\mu$m $\nu_4$ band of CH$_4$ by the \textit{Cassini} Composite
Infrared Spectrometer (CIRS), found to be $(1.6\pm0.5)\times10^{-2}$
\parencite{flasar_05}. Subsequent
observations of the CH$_4$ $\nu_4$ band were used to probe Titan's atmospheric
temperature with CIRS throughout the
\textit{Cassini} mission by holding the GCMS
CH$_4$ profile as latitudinally invariant. However, studies
with the Keck II Near Infrared Spectrometer (NIRSPEC) and the \textit{Cassini}
Visual and Infrared Mapping Spectrometer (VIMS) suggested possible
variations in Titan's atmospheric CH$_4$ 
\parencite{penteado_10a, penteado_10b}. Through the combination of two separate CH$_4$ bands
in focal
planes 1 and 4 of \textit{Cassini}/CIRS, \textcite{lellouch_14} found the
CH$_4$ mole fraction to vary
between $\sim1-1.5\%$ in the lower stratosphere at 12 locations
from $70^{\circ}$ N -- $80^{\circ}$ S during Titan's northern winter ($\sim2005-2010$). This roughly
symmetric distribution of CH$_4$ may persist throughout a Titan year
($\sim29.5$ years), and presents important implications for
photochemical and dynamical models of Titan's atmosphere. 

Monodeuterated methane (CH$_3$D) was tentatively identified on Titan
by \textcite{gillett_75} in observations of the 8.6 $\mu$m band with the Kitt
Peak National Observatory (KPNO), and confirmed
by \textcite{kim_82} in combination with data from the Infrared Telescope Facility (IRTF) and
the \textit{Voyager-1} Infrared Radiometer Interferometer and Spectrometer
(IRIS). Later observations with these facilities 
\parencite{owen_86, debergh_88, coustenis_89b, orton_92, penteado_05}, the
\textit{Infrared Space Observatory} (ISO; \cite{coustenis_03}), and
through both components of the \textit{Cassini-Huygens} mission \parencite{coustenis_07,
  bezard_07, niemann_10, abbas_10, nixon_12} have further constrained
Titan's atmospheric deuterium-to-hydrogen ratio (D/H). With an average
value measured to be $1.36\times10^{-4}$ during the \textit{Cassini}
era (see \cite{nixon_12}, and references therein),
Titan is closer to Earth in terms of atmospheric deuterium enrichment
than the giant planets, leading to possible constraints on the
formation and sustainability of its atmosphere (\cite{mousis_02b};
\cite{bezard_14}, and references therein). 

Due to the dependence on atmospheric temperature for many CH$_4$
observations in the IR, measurements of Titan's CH$_3$D abundance may
provide a second method for determining possible CH$_4$ variations with
latitude, while further constraints on Titan's D/H will help unravel
the history of its formation \parencite{mandt_09}. Previous searches for CH$_3$D in
the interstellar medium using sub-mm facilities were unsuccessful \parencite{pickett_80, womack_96},
or provided a tenuous detection of the $J=1-0$ transition
\parencite{sakai_12}. However, the advent of the Atacama Large
Millimeter/submillimeter Array (ALMA) and Titan's relatively large
atmospheric CH$_3$D content provide a new method to study this molecule
with high spatial and spectral resolution from the ground. Here, we detail the first
definitive detection of CH$_3$D at sub-mm/radio wavelengths. Though these
measurements were produced from short ALMA flux calibration observations of
Titan and do not utilize the full capabilities of the array, they
demonstrate the means by which to study Titan's CH$_4$ distribution
and D/H in the post-\textit{Cassini} era.

\section{Observations} \label{sec:obs}
\begin{figure*}
\centering
\includegraphics[scale=0.95]{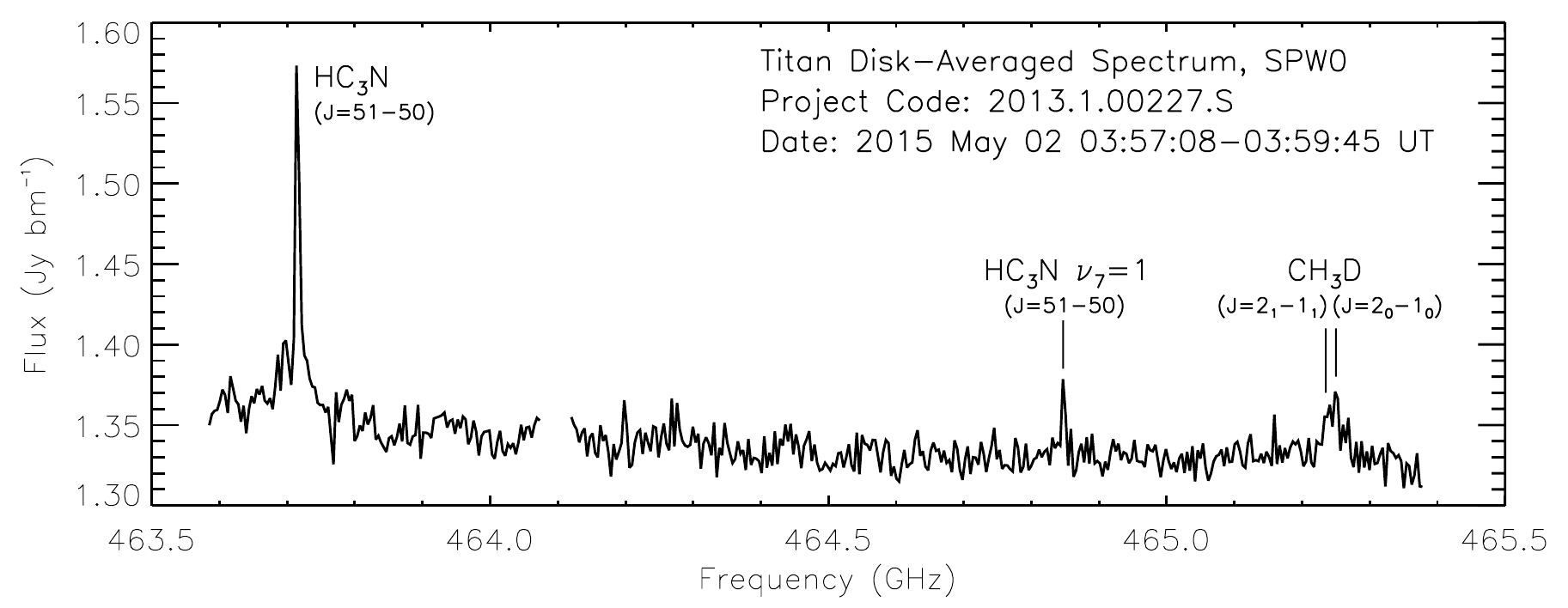}
\caption{The complete disk-averaged spectrum of Titan from
  ALMA observations on 2015 May 02. The spectrum also 
  includes two HC$_3$N lines ($\nu=0$ and $\nu_7=1$) in addition
  to the two CH$_3$D J=2-1 transitions at 465.235 and 465.250
  GHz. Flux at 464.1 GHz has been flagged (removed) during the ALMA data
  reduction process due to terrestrial O$_3$ absorption.}
\label{fig:spec}
\end{figure*}

Titan is often observed as a flux calibration object for ALMA
observations, resulting in many short ($\sim157$ s) observations of
the moon from 2012 onward. As of 2015, ALMA Band 8
($\sim385-500$ GHz) has been available for use, enabling a search
for the $J_K=2_1-1_1$ and $2_0-1_0$ and transitions of CH$_3$D at 465.235 and 465.250 GHz
($\sim0.644$ mm). The
$J=1-0$ and $J=3-2$ transitions at 232.644 GHz (1.289 mm) and
$697.690-697.781$ GHz ($\sim0.430$ mm), respectively, may 
also be observable with ALMA in Bands 6 and 9. However, the former has a lower
line strength than the $J=2-1$ transitions, and was not detected in the
dataset analyzed in \textcite{thelen_18} to model the nearby CO $J=2-1$
transition at 230.538 GHz. As of the time of writing, the only 2 observations of Titan at
frequencies covering the CH$_3$D $J=3-2$ transitions were taken with the lowest
frequency resolution usable with ALMA (31250 kHz), and thus the
CH$_3$D lines were unresolved (if present). Similarly, observations containing
the $J=2-1$ frequencies use low spectral resolution settings or the
Atacama Compact Array (ACA), with the exception of an
  observation on 2015 May 02 at UT 03:57:08, for ALMA Project Code 
$\#$2013.1.00227.S (where Titan was used for flux calibration). We
detected both $J=2-1$ transitions of
CH$_3$D at 4.6$\sigma$ and 5.7$\sigma$ in these data from May 2015, shown in Fig. \ref{fig:spec}. Despite the somewhat coarse spectral
resolution inherent to these data (3904 kHz), we resolve both spectral lines of CH$_3$D,
which are unresolved in additional archival data at 15625
kHz resolution. The analogous $J=2-1$ transitions of the
$^{13}$C-substituted form of CH$_3$D at 464.838 and 464.854 GHz (directly abutting the
HC$_3$N $\nu_7=1$ line, Fig. \ref{fig:spec}) were not detected
in these data, despite the previous detection of this molecule in the
IR \parencite{bezard_07}.

\begin{figure}
\centering
\includegraphics[scale=0.8]{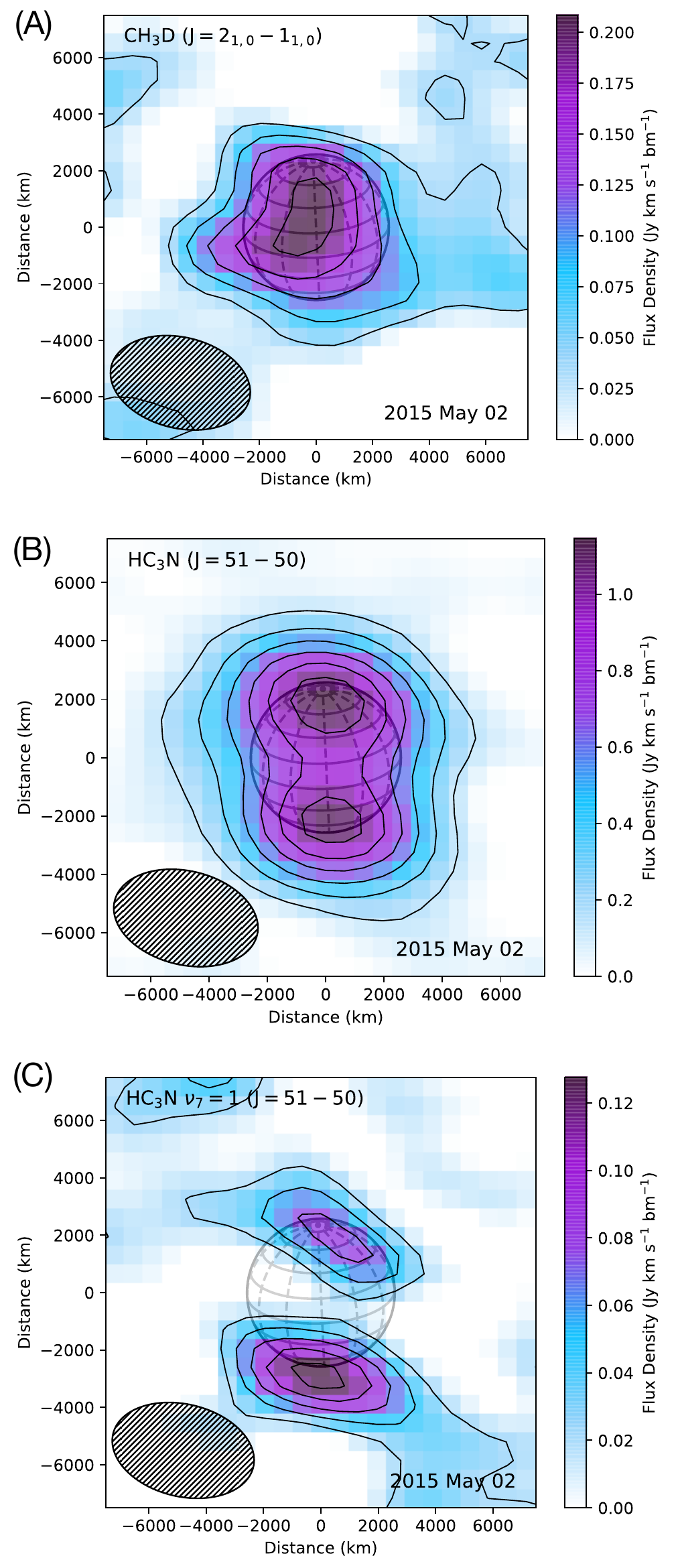}
\caption{Integrated flux maps of the spectral lines shown in
  Fig. \ref{fig:spec}: (A) both CH$_3$D transitions, (B) the ground
  state ($\nu=0$) HC$_3$N
 $J=51-50$ transition, and (C) the corresponding vibrationally excited ($\nu_7=1$) HC$_3$N
  transition. Contour levels are 1$\sigma$ for
  images (A) and (C), and 3$\sigma$ in image (B). The ALMA
  \texttt{clean} beam
  FWHM and orientation are shown as a hatched ellipse. Titan's surface longitude and
  latitude lines are shown in dashed and solid gray lines in
  increments of $30^\circ$ and $22.5^\circ$, respectively.}
\label{fig:images}
\end{figure}

We followed the methodology described in previous works utilizing ALMA
flux calibration data of Titan (e.g. \cite{cordiner_14, lai_17, thelen_18, thelen_19})
to obtain disk-averaged spectra and calibrated image cubes. Using the
Common Astronomy Software Applications (CASA) package version 5.1.2,
we re-ran the ALMA data calibration scripts available with raw data from
the ALMA Science Archive\footnote{http://almascience.nrao.edu} to
ensure that spectral lines in Titan's atmosphere are not flagged (such
as HC$_3$N and CO), and to include the updated flux model for Titan
(see ALMA Memo
$\#$594\footnote{http://library.nrao.edu/public/memos/alma/main/memo594.pdf}). Imaging
was completed using the \texttt{Hogb{\"o}m clean} algorithm with a
pixel size = 
$0.1\times0.1''$ and a flux threshold of 10 mJy -- roughly twice the noise level. The beam
full width at half maximum (FWHM) = $ 0.767\times0.491''$ for this observation, which is
comparable to Titan's angular size on the sky ($\sim0.7-1.0''$, depending
on distance and the inclusion of Titan's substantial
atmosphere). Integrated flux maps of the CH$_3$D lines and both
nearby HC$_3$N transitions are shown in Fig. \ref{fig:images}. Due to
the relatively low signal-to-noise ratio of the CH$_3$D and HC$_3$N
$\nu_7=1$ lines in
this observation and the large beam size compared to Titan's disk,
these data are unsuitable for nuanced interpretation of latitudinal 
variations in Titan's atmosphere. Despite this, the spatial
distribution of both HC$_3$N maps are
consistent with contemporaneous datasets analyzed by \textcite{cordiner_17a}
and \textcite{thelen_19}, and the discrepancies in flux found between $\nu=0$ and
vibrationally excited HC$_3$N lines discussed in \textcite{cordiner_18}. 

\section{Radiative Transfer Modeling} \label{sec:rad}
We extracted a disk-averaged spectrum of Titan by defining a mask of pixels containing 
at least $90\%$ of Titan's continuum flux, as in \textcite{lai_17}. Using the method in
previous studies (e.g. \cite{molter_16, thelen_18}) to extract flux from within
$2\times$ the ALMA PSF resulted in a dilution of
the disk-averaged CH$_3$D lines in these data. The resulting spectrum is shown in
Fig. \ref{fig:spec}. We then converted the disk-averaged flux density into
radiance units (nW cm$^{-2}$ sr$^{-1}$ / cm$^{-1}$) as described in Appendix A of \textcite{teanby_13} using
24 line of sight emission angles to model the flux from the
full disk of Titan and the atmosphere up to 1200 km above the
surface. The array of field-of-view averaging points increases
  in density towards the limb of Titan to account for limb
  brightening, similar to previous disk-averaged models of Titan's
  atmosphere in the sub-mm \parencite{teanby_13, thelen_18}. Although additional emission angles
  may be required to accurately model higher spatial resolution data
  (e.g. \cite{thelen_19}), summing over 24 annuli was sufficient to
  model the disk-averaged radiance from these data. Radiances were
converted using Titan's distance of
9.038 AU. The sub-observer latitude was $24.55^\circ$ and the Doppler
shift was $-7.5$ km s$^{-1}$; these values were obtained from the JPL
Horizons ephemerides
generator\footnote{https://ssd.jpl.nasa.gov/horizons.cgi}. We then
generated initial radiative transfer models using
the Non-linear optimal Estimator for MultivariatE spectral analySIS
(NEMESIS) software package \parencite{irwin_08} to determine any offsets
in the continuum level. 

Titan's continuum was modeled using collisionally-induced
absorption from CH$_4$, H$_2$, and N$_2$ pairs (including isotopes), with coefficients from
\textcite{borysow_86a, borysow_86b, borysow_86c, borysow_87, borysow_91}; and
  \textcite{borysow_93}. As Titan's continuum in
the sub-mm is largely dependent on the tropopause temperature, we constructed
a disk-averaged temperature profile 
using contemporaneous ALMA measurements from \textcite{thelen_18} in the
stratosphere and above, and an
interpolation of tropospheric temperatures found using data 
from the
\textit{Huygens} Atmospheric Structure Instrument (HASI;
\cite{fulchignoni_05}) and those obtained through \textit{Cassini}
radio occultations \parencite{schinder_12}. This resulted in a small ($\textless5\%$ of the initial radiance), constant
offset between the data and model, rectified by multiplying the data by
a factor of 1.047.
This scaling factor is consistent with
previous studies using ALMA flux calibration data (see
\cite{thelen_19}, and references therein), and is due to minor
discrepancies between our Titan model and the one used in the CASA
reduction package for ALMA data (discussed further in \cite{thelen_18}).

\begin{figure}
\centering
\includegraphics[scale=0.8]{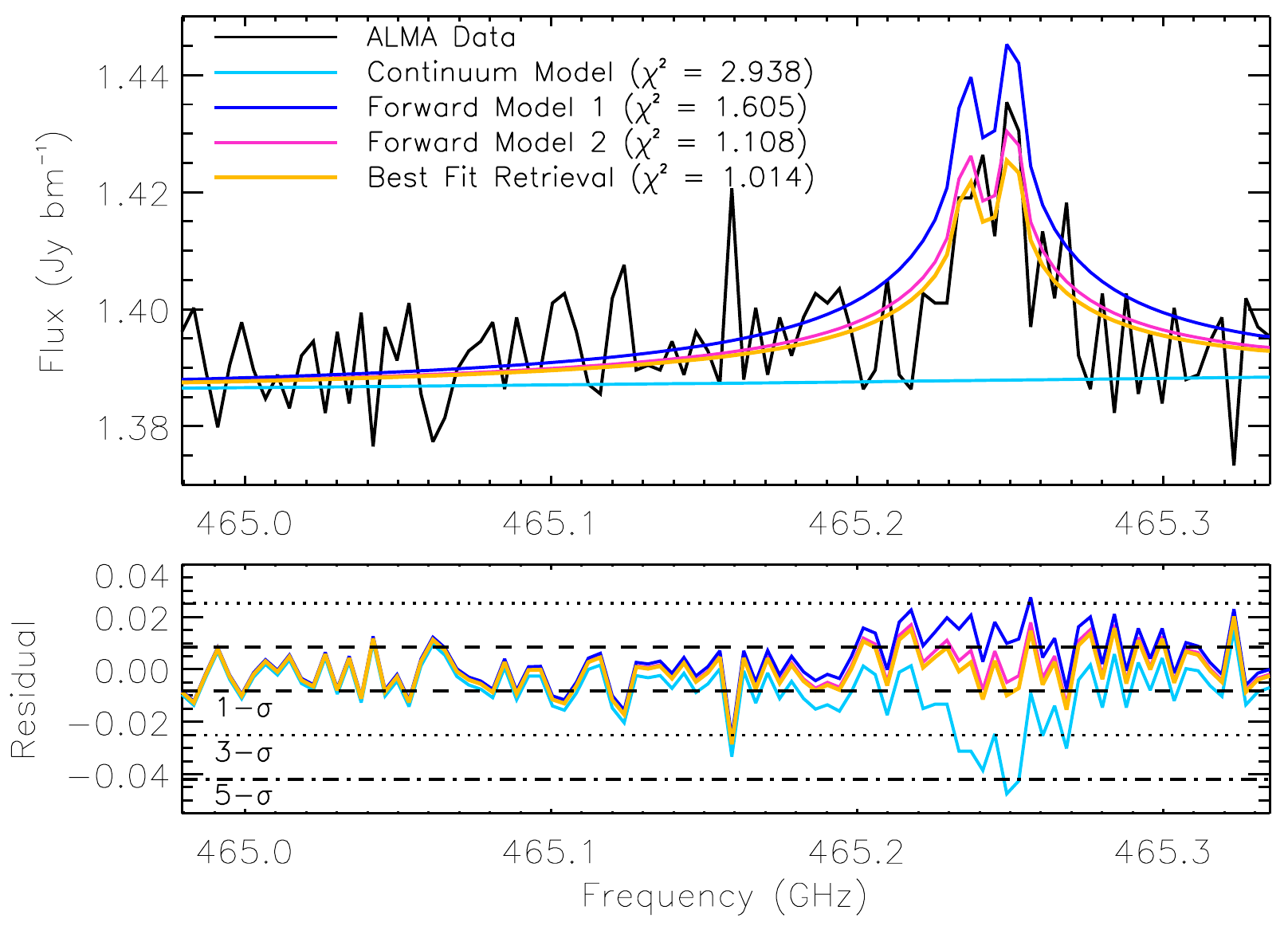}
\caption{(\textit{Top}) Disk-averaged ALMA data (black) with NEMESIS
  radiative transfer models of: Titan's continuum (teal); CH$_3$D spectra corresponding
  to the D/H ratios found by \textcite{nixon_12} and \textcite{coustenis_07}
  (blue and magenta, respectively); the best-fit spectrum found by
  scaling the \textit{a priori} CH$_3$D profiles by NEMESIS
  (gold). Discrepancies between the flux scale here and in
  Fig. \ref{fig:spec} are due to the implementation of a scaling
  factor during modeling. The
  spectral feature at $\sim$465.16 GHz appears to be
  a noise spike, as it does not correspond to any other minor
  species used in our models of Titan's atmosphere. (\textit{Bottom}):
  Residual spectra (model$-$data) for the models in the top panel. $\pm$1, 3, and
  $5\sigma$ thresholds are shown in dashed, dotted, and dash-dot
  lines, respectively.}
\label{fig:models}
\end{figure}

\begin{figure}
\centering
\includegraphics[scale=0.8]{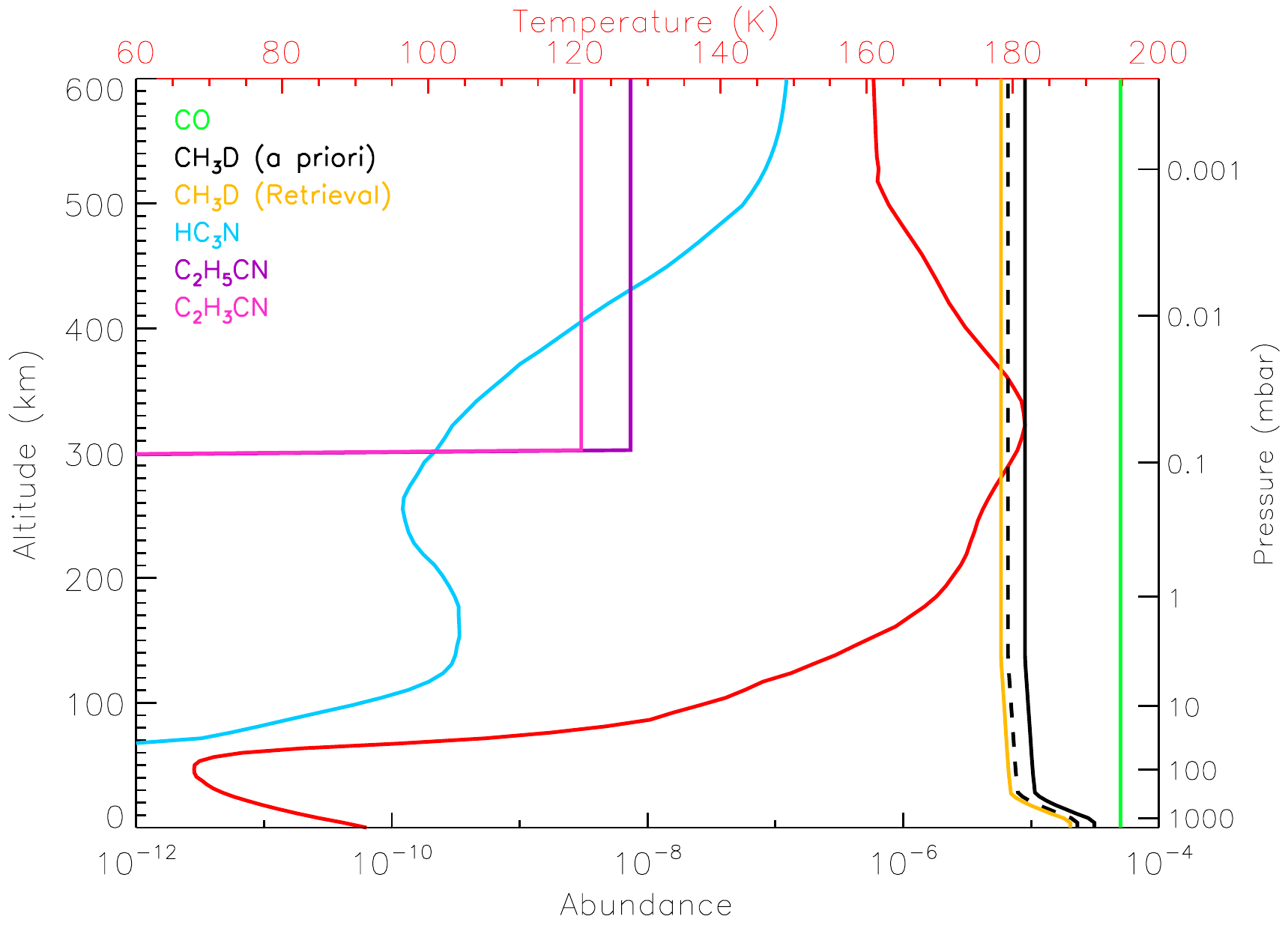}
\caption{Atmosphere profiles used in our NEMESIS radiative
  transfer model: CO (green); HC$_3$N (teal, from
  \cite{thelen_19}); C$_2$H$_3$CN and C$_2$H$_5$CN (pink and
  purple, respectively) from \textcite{lai_17}; temperature (red) from
  \textcite{thelen_18}. The CH$_3$D profiles used as \textit{a
    priori} inputs are shown in black, corresponding to the D/H ratios
  from \textcite{nixon_12} (solid line)  and
  \textcite{coustenis_07} (dashed line). Our best-fit model is shown in
  gold, corresponding to a D/H = $1.033\times10^{-4}$.}
\label{fig:abund}
\end{figure}

To obtain an accurate radiative transfer model of the CH$_3$D lines
emitted in Titan's atmosphere, we ran NEMESIS in line-by-line mode
over a small spectral range, shown in Fig. \ref{fig:models}. CH$_3$D spectral line and partition function 
parameters were obtained from the Cologne
Database for Molecular Spectroscopy (CDMS; \cite{muller_01,
  womack_96, bray_17}). Line
broadening and temperature dependence coefficients were
obtained from the HITRAN 2012 database \parencite{rothman_13}. Our
NEMESIS model also consists of nearby species that may influence the CH$_3$D
line shape and nearby continuum, and an appropriate temperature-pressure profile
for Titan during northern summer. We thus
include N$_2$ and CH$_4$
vertical abundance profiles from \textcite{niemann_10} and
\textcite{teanby_13}; CO with a constant volume mixing
ratio of 50 ppm (see \cite{serigano_16},
and references therein); HC$_3$N, using the 2015 disk-averaged
abundance profile found in
\textcite{thelen_19}. Though nearby emission lines of
C$_2$H$_3$CN and C$_2$H$_5$CN do not strongly affect the line shape of
the CH$_3$D lines and are undetected in the modeled spectral range of
these data, these
species were included for completeness using abundance profiles from
\textcite{lai_17}. The vertical temperature and abundance profiles
included in our model are shown in Fig. \ref{fig:abund}.

Due to the somewhat coarse spectral resolution settings of this
observation and the relatively low signal-to-noise ratio of both CH$_3$D lines (compared to the
nearby HC$_3$N line, for example), we opted to find a best-fit
spectrum using a
simple scaling model (i.e. a constant multiplicative factor) of an
\textit{a priori} CH$_3$D profile. All other modeling parameters, primarily the temperature and additional
gas abundance profiles, were held constant, and the CH$_3$D profile
was set to retain the initial shape of the \textit{a priori} input (that is,
NEMESIS did not retrieve a continuous vertical profile). Our initial CH$_3$D profiles were
found by multiplying \textit{in situ} CH$_4$ data obtained with 
the \textit{Huygens} probe \parencite{niemann_10} by various D/H
ratios. These include measurements made during the
  \textit{Cassini} era from previous
ground-based studies \parencite{penteado_05, debergh_12}, the
\textit{Huygens} GCMS 
\parencite{niemann_10}, and with \textit{Cassini}/CIRS \parencite{coustenis_07, bezard_07,
  abbas_10, nixon_12}, and cover a range of D/H ratios from
$(1.13-1.59)\times10^{-4}$. Spectral models using the D/H ratios =
$1.17\times10^{-4}$ and $1.59\times10^{-4}$ from \textcite{coustenis_07}
and \textcite{nixon_12}, respectively, are shown in Fig. \ref{fig:models}
(magenta and blue spectra); the corresponding CH$_3$D abundance
profiles are shown in Fig. \ref{fig:abund} (black lines). Models using
all \textit{a priori} abundance profiles converge on a single best-fit
spectrum shown in Fig. \ref{fig:models} (gold), with scaling factors
producing a common CH$_3$D profile shown in Fig. \ref{fig:abund} (gold
line). Our retrievals were most sensitive between $\sim100-200$ km,
where we find the CH$_3$D abundance to decrease from
($6.455-6.157$)$\times10^{-6}$.

\section{Discussion and Conclusions} \label{sec:disc}
Using archival ALMA flux calibration data of Titan from 2015, we have produced
the first definitive detection of CH$_3$D in the sub-mm. While
integrated flux maps of the ground ($\nu=0$) and vibrationally excited ($\nu_7=1$)
HC$_3$N transitions produced from the same image cube
(Fig. \ref{fig:images}B, C) show spatial
distributions consistent with previous studies of Titan with ALMA
\parencite{cordiner_17a, cordiner_18, thelen_19}, the integrated flux map
of CH$_3$D (Fig. \ref{fig:images}A) shows a lack of significant 
variation between the poles (less than $\sim30\%$). Though this is
consistent with a constant CH$_3$D profile (and that of its parent
molecule, CH$_4$) with latitude, the low spatial
resolution and S/N of the CH$_3$D lines prohibit the interpretation of
possible latitudinal variations seen in the higher 
resolution data acquired with \textit{Cassini}/VIMS and CIRS
\parencite{penteado_10b, lellouch_14}. We thus measured the
disk-averaged abundance of CH$_3$D using the NEMESIS
radiative transfer code to be a constant value = $6.157\times10^{-6}$ above $\sim130$
km, where our measurements are most sensitive. When taken with the CH$_4$ profile found by the \textit{Huygens}
GCMS \parencite{niemann_10}, our CH$_3$D abundance yields a D/H =
($1.033\pm0.081)\times10^{-4}$. 

The D/H ratio found above is within the errors of previous
measurements from ISO, IRTF, KPNO, and the \textit{Cassini-Huygens} mission
\parencite{coustenis_03, penteado_05, coustenis_07, niemann_10,
  debergh_12}, though generally lower than the average of measurements
made during the \textit{Cassini} era ($1.36\times10^{-4}$;
\cite{nixon_12}), and from \textit{Voyager-1} ($1.5\times10^{-4}$; \cite{coustenis_89b}). The addition of our measurement results in a new
weighted mean D/H = ($1.203^{+0.057}_{-0.054})\times10^{-4}$ for
observations made during the \textit{Cassini/Huygens} mission, or
utilizing GCMS results. The error bars on our measurement are
relative, and only reflect the retrieval errors of our best-fit
synthetic CH$_3$D spectrum. As we cannot simultaneously detect CH$_4$
with ALMA, any D/H measurements rely on previous data obtained in the
IR by ISO, \textit{Voyager-1, Cassini}, various ground-based
facilities, or \textit{in situ} data from the \textit{Huygens}
probe. As the GCMS abundance profile of CH$_4$ was determined at the
probe's landing site ($\sim10^{\circ}$ S), disk-averaged
measurements of Titan with ALMA may not yield a completely comparable
D/H to previous, self-consistent measurements with the
\textit{Cassini-Huygens} mission. As shown during the
  \textit{Voyager-1} mission and with the ISO \parencite{coustenis_89b,
    coustenis_03}, the measurement of Titan's D/H ratio is often not
  straightforward without a well constrained measurement of
  CH$_4$. These early results were determined using CH4 abundances =
  1.8--1.9$\%$, which would significantly
  lower our D/H measurement ($\sim8.1\times10^{-5}$) if used instead of the
  \textit{Huygens}/GCMS results.
Further, as there is only one dataset on the ALMA archive
(at the time of writing) with the spectral settings and resolution capable of
resolving both $J=2-1$ transitions of CH$_3$D, we were unable to stack
measurement sets to produce a higher S/N detection. 

As such, the D/H
value we report should be taken with caution, as there are many
potential sources of error that are difficult to quantify in
relatively low spatial resolution ALMA data -- including
variations in temperature and CH$_4$ abundance with
latitude. However, despite our relatively low disk-averaged value of D/H 
  compared to \textit{Cassini-Huygens} measurements, it remains larger than the D/H values
  derived from CH$_4$ abundances for Jupiter ($2.2\times10^{-5}$,
  \cite{lellouch_01}) and Saturn ($1.6\times10^{-5}$,
  \cite{fletcher_09}) by almost an order of magnitude and is more
  in line with terrestrial D/H measurements in Earth's oceans
  (D/H for Vienna Standard Mean Ocean Water is $\sim1.56\times10^{-4}$; see, e.g. \cite{alexander_12}). Thus, our
  measurement is consistent with previous observations and models that 
  indicate Titan's CH$_4$ reservoir did not evolve from a Saturnian D/H
  value (e.g. through photochemistry or serpentinization reactions within
  Titan's interior; \cite{cordier_08, mousis_09b}). Instead,
  Titan was possibly formed from an aggregate of planetesimals that
  were initially enriched in deuterium from interstellar CH$_4$. The
  isotopic exchange between infalling gas phase CH$_4$ and H$_2$ in the
  solar nebula eventually resulted in an enhanced D/H value, which was
  then preserved as CH$_4$ condensed or was trapped within water-ice
  clathrates in the Saturnian subnebula \parencite{mousis_02a,
    mousis_02b}. Titan's D/H value was then modified
  through photochemistry, diffusion, atmospheric escape, and the outgassing of CH$_4$ from the
  interior \parencite{cordier_08, mandt_09}, finally resulting in near an order of
  magnitude greater deuterium enrichment compared to Saturn as we
  find here.

Despite the uncertainties of our D/H measurement, the detection of
CH$_3$D in the sub-mm allows for independent
observations of the spatial variation in atmospheric CH$_4$ without
the large dependence on simultaneous temperature measurements (as are
required in many IR observations). As noted
above, the lack of ALMA flux calibration observations covering the
$J=2-1$ and the adjacent transitions of CH$_3$D presents a need for
targeted study of Titan with longer integration times and adequate
spectral resolution. With ALMA's high spatial resolution capabilities, further studies of
CH$_3$D may enable mapping of CH$_4$ down to 100's of km allowing
for a follow-up to the study by \textcite{lellouch_14}; higher S/N data
will also result in CH$_3$D abundance measurements from a larger range
of altitudes. Finally,
measuring Titan's CH$_3$D abundance near the equator allows for a
more constrained D/H when
determined using data from the \textit{Huygens} probe.

\section{Acknowledgments}
AET and MAC were funded by the National Science Foundation Grant $\#$AST-1616306. 
CAN and MAC received funding from NASA's Solar System Observations
Program. CAN was supported by the NASA Astrobiology Institute. PGJI acknowledges the support of the UK Science and Technology Facilities Council.

This paper makes use of the following ALMA data:
ADS/JAO.ALMA$\#$2013.1.00227.S, 2013.1.01010.S, and 2012.1.00688.S. ALMA
is a partnership of ESO
(representing its member states), NSF (USA) and NINS (Japan), together
with NRC (Canada) and NSC and ASIAA (Taiwan) and KASI (Republic of
Korea), in cooperation with the Republic of Chile. The Joint ALMA
Observatory is operated by ESO, AUI/NRAO and NAOJ. The National Radio
Astronomy Observatory is a facility of the National Science Foundation
operated under cooperative agreement by Associated Universities, Inc.

\printbibliography[title={References}]

\end{document}